\documentstyle[11pt]{article}
\begin{document}

\centerline{\bf Quantum dynamical theory for squeezing the output of}
\centerline{\bf a Bose-Einstein condensate}

\bigskip

\centerline{Hui Jing$^a$, Jing-Ling Chen$^a$, Mo-Lin Ge$^{a,b}$}
\bigskip\

\centerline{$^a$Theoretical Physics Division, Nankai Institute of
Mathematics,  }
\centerline{Nankai University, Tianjin 300071, People's Republic of China}
\centerline{ Email:   hjing2000@eyou.com}
\centerline{ Email:   jinglingchen@eyou.com}
\centerline{$^b$Center for Advanced Study, Tsinghua University,}
\centerline{ Beijing 100084, People's Republic of China}

\bigskip\

\begin{center}
{\bf Abstract}
\end{center}

\noindent

 A linear quantum dynamical theory for squeezing the output of the
 trapped Bose-Einstein condensate is presented with the Bogoliubov
 approximation. We observe that the non-classical properties, such
 as sub-Poisson distribution and quadrature squeezing effect,
 mutually oscillate between the quantum states of the applied optical
 field and the resulting atom laser beam with time. In particular,
 it is shown that an initially squeezed optical field will lead to
 squeezing in the outcoupled atomic beam at later times.

\bigskip 

{\bf PACS number(s): 03.75.Fi}

\newpage\ 
{\small

\noindent {\bf I. INTRODUCTION}

\bigskip

\noindent
Since the first observations of Bose-Einstein condensation in an atomic gas
in 1995[1,2], there have been many interests in creating an atom laser and
exploring its novel properties. In 1997[3], the MIT-group first realized a
pulsed atom laser, by using a RF pulse to transfer the initially trapped
condensate into the untrapped state. Later, successive experimental
achievements in the design and amplification of atom laser[4-7] were
obtained
and stimulated amounts of theoretical works in both the output coupling and
the properties of atom laser[8-10].

Recently, much attention were also paid to the problem of nonlinear atomic
optics. Deng et al., for example, realized the four matter-wave mixing in
their remarkable experiment by applying the optical technique of Bragg
diffraction to the condenste[11]. Most recently, Moore et al. pointed out
the
possibility to realize the optical control on the quantum statistics of the
output matter wave in the framework of nonlinear atomic optics[12]. 

The aim of this paper is to investigate the possibility to produce a
squeezed atom laser from the trapped condensate, as the analogy of the
squeezed light which is now available in laboratory[13]. The paper is
organized as the following. In Sec. II, we present a model for squeezing
the output of the trapped atoms with a many-boson system of two
states, trapped state and untrapped state, with linear coupling. In the
Bogoliubov approximation, the solutions of this system is also derived.
Based on these solutions, the non-classical properties such as sub-Poisson
distribution and quadrature squeezing effect are investigated in Sec. III,
which leads to an interesting oscillation behavior of the quantum statistics
between the coupling light field and the output atomic field. Finally,
Sec. IV is a summary and outlook.

\bigskip

\noindent {\bf II. MODEL AND SOLUTIONS}

\bigskip

\noindent
For simplicity, we shall assume that the atoms have two states, $|1\rangle $
and $|2\rangle $, with the initial condensation occurring in the trapped
state
$|1\rangle $. State  $|2\rangle $, which has different trapping properties
and
is typically unconfined by the magnetic trap, is coupled to  $|1\rangle $ by
a one-mode squeezed optical field tuned near the
$|1\rangle \rightarrow |2\rangle $ transition. The interaction of the field
may thus generate condensate in state $|2\rangle$, from an initial
condensate
which is entirely in state $|1\rangle$. The Hamiltonian of this dynamical
model
with linear coupling interaction can be written as[14] $(\hbar=1)$

\begin{equation}
H= \omega _0 b_2^{\dag }b_2+ \omega _a a^{\dag }a
+ \omega _R^{'}  ( a b_1 b_2^{\dag}+a^{\dag} b_1^{\dag} b_2 )
\end{equation}
in terms of the creation and annihilation operators,
$b_1^{\dag },$ $b_2^{\dag }$, $b_1$ and $b_2$, of bosonic atoms for the
magnetically trapped state $|1\rangle $ and the untrapped state $|2\rangle $
with level difference $\omega _0$, $a^{\dag}$ and $a$ is the creation and
annihilation operators of the optical field with frequency $\omega_a$. Here
$\omega_R^{'}=\sqrt{\omega_a/  2 \varepsilon _0 V}$,
$V$ is the effective mode volume and $\varepsilon _0$ is the vacuum
permittivity. The nonlinear interaction between the atoms and the quantized
motion of atomic center of mass in the trapped state by an inhomogeneous
magnetic field has been ignored, which is the main simplification of our
model.

  We suppose the initial state of the system is theoretically described as
$|\psi(0)\rangle =|\alpha\rangle _1\otimes
|\Phi (0)\rangle _s$ with
$|\Phi(0)\rangle_s =|0\rangle _2\otimes |\xi \rangle $.
Here $|\alpha \rangle _1$ is a Glauber coherent state of the operator
 $b_1$ characterizing the condensed atoms in the trapped state $|1\rangle $,
namely, $b_1|\alpha \rangle=\sqrt{N_c}e^{-i\theta}|\alpha\rangle$;
$|0\rangle _2$ represents that the initial untrapped state $|2\rangle $ is a
vacuum state since there is no occupying atoms in it;
and the initial state of the input optical field is the squeezed state[15]:
$|\xi \rangle=S(\xi)|m\rangle$, where the squeezed operator
$S(\xi)=\exp[\xi (a^{\dag})^2-{\xi}^*a^2]$ with
$\xi=\frac{1}{r}\exp(-2i\phi)$,
representing a unitary transformation on the coherent state $|m\rangle$. 
In the Bogoliubov approximation[16], we can ignore the slow change of the
large number $N_c$ of the condensed atoms in the trap, which means that the
operators $b_1$, $b_1^{\dag }$ can be replaced with a $c$-number
$\sqrt{N_c}$.
Therefore, the condensed component initially in a coherent state
$|\alpha \rangle_1$ remains in such a state while another component
$|\Phi (0)\rangle _s$ is governed by the Bogoliubov approximate
Hamiltonian[14].

     The Heisenberg equations about the operators $b_2$ and $a$ are

\begin{equation}
i \frac{\partial}{\partial t} b_2= \omega_0 b_2+\omega_R^{'}a b_1, \;\;
i \frac{\partial}{\partial t} a= \omega_a a+\omega_R^{'} b^{\dag}_1 b_2.
\end{equation}
After taking the average value with respect to the coherent state
$|\alpha \rangle$, the above equations can be written as

\begin{equation}
i \frac{\partial}{\partial t}
\left (
  \begin{array}{c}
    b(t)\\
    a(t)
   \end{array}
\right )
=
\left (
  \begin{array}{cc}
  \omega_0  & \omega_R e^{-i\theta} \\
  \omega_R e^{i\theta} & \omega_a
  \end{array}
\right )
\left (
  \begin{array}{c}
    b(t)\\
    a(t)
   \end{array}
\right ),
\end{equation}
where $\omega_R=\omega_R^{'} \sqrt{N_c}$ and $b_2$ was rewritten as $b$. By
applying the technique of diagonalizing the coefficient matrix, we can
obtain
the exact solutions of the coupling equations as follows:

\begin{equation}
\left (
  \begin{array}{c}
    b(t)\\
    a(t)
   \end{array}
\right )
=
\left (
  \begin{array}{cc}
  \lambda_-(t) & -i \eta (t) e^{-i\theta} \\
  -i \eta (t) e^{i\theta} & \lambda_+(t)
  \end{array}
\right )
\left (
  \begin{array}{c}
    b(0)\\
    a(0)
   \end{array}
\right )
e^{-i\frac{1}{2}(\omega_0+\omega_a)t},
\end{equation}
where
$\lambda_{\pm}(t)=\cos(I(\varphi)t)\pm i \sin(\varphi) \sin (I(\varphi)t)$,
$\eta (t)= \cos\varphi \sin(I(\varphi)t)$, and $\varphi$, $I(\varphi)$
are defined
as $\omega_0 - \omega_a = 2\omega_R \tan(\varphi)$,
 $I(\varphi)=\omega_R /\cos(\varphi)$ and $\varphi$ is an arbitrary
constant.
   Now we consider the special case with a resonance frequency, namely, 
$\omega_a=\omega_0=\omega$ (or $\varphi=2n\pi$, $n=${\rm integer})
and $I(\varphi)=\omega_R$, the eq.(4) can then be simplified as

\begin{equation}
\left (
  \begin{array}{c}
    b(t)\\
    a(t)
   \end{array}
\right )
=
\left (
  \begin{array}{cc}
  \cos(\omega_R t) & -i \sin(\omega_R t)e^{-i\theta} \\
  -i \sin(\omega_R t)e^{i\theta} & \cos(\omega_R t)
  \end{array}
\right )
\left (
  \begin{array}{c}
    b(0)\\
    a(0)
   \end{array}
\right )
e^{-i\omega t}.
\end{equation}

  Obviously, for the evolution times satisfying $\cos(\omega_R t_0)=0$ or
   $\omega_R t_0=(n+1/2)\pi$, we have
 
\begin{equation}
b(t_0)=-i (-1)^n a(0) e^{-i(\omega t_0+\theta)}, \;\;
a(t_0)=-i (-1)^n b(0) e^{-i(\omega t_0-\theta)},
\end{equation}
where the parameter $\theta$ will not affect our main results, as it can be
seen below. In particular, eq.(6) leads to

\begin{equation}
< b^{\dag} (t_0) b(t_0)>=< a^{\dag} (0) a(0)>,
\end{equation}
which indicates a complete $quantum$ $conversion$ of the statistical
properties
between the input photons and the output atoms.

\bigskip

\noindent {\bf III. NON-CLASSICAL PROPERTIES OF THE OUTPUT ATOMIC FIELD}

\bigskip

\noindent
Using the solutions obtained above, we can calculate
the average numbers and the fluctuations of the out-state photons as well as
the output atoms. For the out-state photons, the final results are
$$< N_a(t)>_s=<\Phi(0)|  a^{\dag} (t) a(t)|\Phi(0)>$$
\begin{equation}
=\{m^* m \alpha_1 +[(m^*)^2 e^{-2i\phi}+m^2 e^{2i\phi}]\alpha_2
+\sinh^2 r\} \cos^2 (\omega_R t),
\end{equation}
and

$$< \Delta N_a^2 (t)>_s=< N_a^2 (t)>_s -< N_a (t)>_s^2$$

$$=\{ |m|^2 \alpha_1^2 +2 \alpha_2^2( 2|m|^2+1)
+2\alpha_1 \alpha_2 [(m^*)^2 e^{-2i\phi}+m^2 e^{2i\phi}]
\} \cos^4(\omega_R t)$$

\begin{equation}
+\{ \alpha_1 |m|^2 + \sinh^2r + [(m^*)^2 e^{-2i\phi}+m^2 e^{2i\phi}]\alpha_2
\} \sin^2(\omega_R t) \cos^2(\omega_R t),
\end{equation}
where $\alpha_1= \sinh^2 r+ \cosh^2 r$ and $\alpha_2= \sinh r \cosh r$.
Similarly, the fluctuation of the output atoms is
$$< \Delta N_b^2 (t)>_s=< N_b^2 (t)>_s -< N_b (t)>_s^2$$
$$=\{ |m|^2 \alpha_1^2 +2 \alpha_2^2( 2|m|^2+1)
+2\alpha_1 \alpha_2 [(m^*)^2 e^{-2i\phi}+m^2 e^{2i\phi}]
\} \sin^4(\omega_R t)$$
\begin{equation}
+\{ \alpha_1 |m|^2 + \sinh^2 r+ [(m^*)^2 e^{-2i\phi}+m^2 e^{2i\phi}]\alpha_2
\} \sin^2(\omega_R t) \cos^2(\omega_R t).
\end{equation}

 Obviously, if the squeezed angle is chosen as $\phi =0$ and
$m\in {\rm R(real \; number)}$, we then obtain
$$< \Delta N_a^2 (t)>_s=[ m^2 (\alpha_1 +2 \alpha_2)^2+ 2\alpha_2^2 ]
 \cos^4(\omega_R t)
$$
\begin{equation}
+[ \sinh^2 r+(\alpha_1+2\alpha_2) m^2] \sin^2(\omega_R t) \cos^2(\omega_R
t),
\end{equation}
$$< \Delta N_b^2 (t)>_s=[ m^2 (\alpha_1 +2 \alpha_2)^2+ 2\alpha_2^2]
 \sin^4(\omega_R t)
$$
\begin{equation}
+[ \sinh^2 r+(\alpha_1+2\alpha_2) m^2 ] \sin^2(\omega_R t) \cos^2(\omega_R
t).
\end{equation}
In particular, if the initial optical field is in a vacuum-squeezed state
($m=0$), the results can be written in the following simple form:
\begin{equation}
< N_a(t)>_s^0=\sinh^2 r \cos^2(\omega_R t), \;\;
< N_a^2(t)>_s^0=(2\alpha_2+ \sinh^4 r) \cos^4(\omega_R t),
\end{equation}
and 
\begin{equation}
< \Delta N_a^2(t)>_s^0=\sqrt{2}\sinh r \cos^4(\omega_R t), \;\;
< \Delta N_b^2(t)>_s^0=\sqrt{2}\sinh r \cosh r \sin^4(\omega_R t).
\end{equation}
Now we suppose that the evolution times satisfy the following conditions
\begin{equation}
\cos(\omega_R t)=0 ,\;\;  {\rm or} \;\;  \omega_Rt=(n+1/2)\pi,
~~(n=0,1,2,...)
\end{equation}
then eqs.(11) and (12) become

\begin{equation}
< \Delta N_a^2(t)>_s^0=0,\;\;
< \Delta N_b^2(t)>_s^0=m^2(\alpha_1+2\alpha_2)^2+2\alpha_2^2.
\end{equation}
Obviously, even if we set $m=0$, we have:
$< \Delta N_b(t)>_s^0=\sqrt{2} \sinh r \cosh r \neq 0$,
which characterizes the existence of the squeezing effect for the output
atomic field.

It is well-known that, to decide the statistical properties of a
quantum field, we can define a $Q$ parameter as[15]:

\begin{equation}
Q_a^s (t) =\frac{< \Delta N_a^2(t)>_s}{< N_a(t)>_s} -1
\left \{
\begin{array}{ll}
<1 : & {\rm Sub-Poisson\;  distribution,  } \\
=0 : & {\rm Poisson\; distribution,  } \\
>0 : & {\rm Super-Poisson\;  distribution.   }
\end{array}
\right.
\end{equation}
From eqs.(8), (11) and (12), the $Q$ parameters of the out-state optical
field and the output atomic field can be derived as ($\phi =0, m\in {\rm
R}$)

\begin{equation}
\left (
  \begin{array}{c}
    Q_a^s(t)\\
    Q_b^s(t)
   \end{array}
\right )
=
[\frac{m^2(\alpha_1+2\alpha_2)^2+2\alpha_2}
{m^2(\alpha_1+2\alpha_2)+\sinh^2 r}-1]
\left (
  \begin{array}{c}
   \cos^2(\omega_R t) \\
    \sin^2(\omega_R t)
   \end{array}
\right ),
\end{equation}
which yields an interesting result for $m=0$:
\begin{equation}
\left (
  \begin{array}{c}
    Q_a^s(t)\\
    Q_b^s(t)
   \end{array}
\right )
=
\alpha_1
\left (
  \begin{array}{c}
    \cos^2(\omega_R t) \\
    \sin^2(\omega_R t)
   \end{array}
\right ).
\end{equation}
In the initial state ($t=0$), $Q_a^s(t)>0,~~Q_b^s(t)=0$, which means
the initial state of the optical field is a squeezed state and the
initial state of the atom field is a vacuum state, as they should be.
When the evolution time $t_0$ satisfies $\cos(\omega_R t_0)=0$,
then $Q_a^s(t)=0,~~Q_b^s(t)>0$, which means the initial squeezed optical
field transforms into a coherent state as well as the initial coherent atom
field is now squeezed. It is this interesting periodically
oscillating behavior which holds the promise to be observed in the
laboratory.

At last, we proceed to calculate the quadrature squeezing of output atomic
field. According to Ref.[15], the field quadratures $X_{1a}$, $X_{1b}$,
$X_{2a}$ and $X_{2b}$ are defined as

\begin{equation}
X_{1a}=\frac{1}{2}(a+a^{\dag}), \;\; X_{2a}=\frac{1}{2i}(a-a^{\dag}),\;\;
X_{1b}=\frac{1}{2}(b+b^{\dag}), \;\; X_{2b}=\frac{1}{2i}(b-b^{\dag}).
\end{equation}
Following Bruzek et al.[17], we introduce the squeezed coefficients

\begin{equation}
S_i=\frac{<(\Delta X_i)^2>-\frac{1}{2}|<[X_1,X_2]>}
         {\frac{1}{2}|<[X_1,X_2]>|}, \;\; i=1,2
\end{equation}
one then gets
$$S_{1b}(t)=2<N_b(t)>+2 {\rm Re}<b^2(t)>-({\rm Re}<b(t)>)^2,$$
\begin{equation}
S_{2b}(t)=2<N_b(t)>-2 {\rm Re}<b^2(t)>-({\rm Im}<b(t)>)^2.
\end{equation}
On account of ($\phi=0, m=0$)
\begin{equation}
<b^2(t)>_s=-\sinh r \cosh r e^{-2i\omega t} \sin^2 (\omega_R t),\;\;
<b^{\dag} b(t)>_s=\sinh^2 r \cosh^2 r \sin^2 (\omega_R t),
\end{equation}
and $<b(t)>_s=0$, we can finally obtain
$$S_{1b}(t)=2\sinh r \{\sinh r -\cosh r \cos[2(\omega t+\theta)]\}
\sin^2(\omega_R t),$$
\begin{equation}
S_{2b}(t)=2\sinh r \{\sinh r +\cosh r \cos[2(\omega t+\theta)] \}
\sin^2(\omega_R t).
\end{equation}
Obviously, for the initial state of the atomic field, eq.(24) yields
$S_{1b}(0)=S_{2b}(0)=0$,
which means there is no squeezing, as it should be.

After a period of evolution time, namely, $\omega t+\theta=n \pi$, but
$\omega_R t \neq n \pi$, we can get the following results for the output
atomic field

\begin{equation}
S_{1b}(t)=-2\sinh r e^{-r} \sin^2 (\omega_R t)<0,\;\;
S_{2b}(t)=2\sinh r e^{-r} \sin^2 (\omega_R t)>0,
\end{equation}
which just means that the quadrature component $X_{1b}$ is squeezed.

However, after another period of evolution time, namely,
$\omega t+\theta=(n+1/2)\pi$, but $\omega_R t \neq n \pi$, it will become

\begin{equation}
S_{1b}(t)=2\sinh r e^{-r} \sin^2 (\omega_R t)>0,\;\;
S_{2b}(t)=-2\sinh r e^{-r} \sin^2 (\omega_R t)<0,
\end{equation}
which means that the squeezing effect transfers to $X_{2b}$ component.
In the same
way, we can also calculate the squeezed coefficients for the optical field,
which shows a similar behavior as the atomic field.

\bigskip

\noindent {\bf IV. CONCLUSION AND OUTLOOK}

\bigskip

 In this paper, we have theoretically presented a model for squeezing
the output of the trapped condensed atoms with a many-boson system of two
states with linear coupling. In the Bogoliubov approximation, its solutions
for the many-body problem show that the non-classical properties, such as
sub-Poisson distribution and quadrature squeezing effect, mutually oscillate
between the quantum states of the applied optical
field and the resulting atom laser beam with time. Hence after some
period of coupling interaction, the initially squeezed light will transform
into a coherent state while the output atomic field is in a squeezed state,
which means a squeezed output of atomic beam in the propagating mode.

 The availability of a squeezed atom laser would certainly be useful in
 future applications of cold atoms and our investigation provides
 a beginning point towards its realization. However, our model does not
 include atom-atom interactions and therefore only works when the atomic
 field is very dilute. Future work will involve generalizing this squeezed
 atom laser model to include the influence of atomic interactions, the
 effects of the trapping field and the shape of the condensate. Furthermore,
 it is also important to analyze in detail the requirements for the actual
 experimental realization (which would probably require an optical cavity)
 and the conditions for the relevant parameters, which may deserve further
 works in the future.

\noindent

\newpage 

\noindent
{\bf ACKNOWLEDGMENT}
\bigskip

This work was partially supported by the National Natural Science
Foundation of China.


\begin{thebibliography}{99}

\bibitem{1}  M. H. Anderson, J. R. Ensher, M. R. Matthews, C. E. Wieman
             and E. A. Cornell, Science {\bf 269}, 198 (1995).
\bibitem{2}  K. B. Davis, M.-O. Mewes, M. R. Andrews, N. J. van Druten,
             D. S. Durfee, D. M. Kurn and W. Ketterle, Phys. Rev. Lett.
             {\bf 75}, 3969 (1995).
\bibitem{3}  M.-O Mewes, M. R. Andrews, D. M. Kurn, D. S. Durfee,
             C. G. Townsend and W. Ketterle, Phys. Rev. Lett {\bf 78},
             582 (1997).
\bibitem{4}  B. P. Anderson and M. A. Kasevich, Science {\bf 282},
             1686 (1998).
\bibitem{5}  E. W. Hagley, L. Deng, M. Kozuma, J. Wen, K. Helmerson,
             S. L. Rolston and W. D. Phillips, Science {\bf 283}, 1706
             (1999).
\bibitem{6}  I. Bloch, T. W. H{\"a}nsch and T. Esslinger, Phys. Rev. Lett
             {\bf 82}, 3008 (1999).
\bibitem{7}  M. Kozuma, Y. Suzuki, Y. Torii, T. Sugiura, T. Kuga,
             E. W. Hagley and L. Deng, Science {\bf 286}, 2309 (1999).
\bibitem{8}  G. M. Moy and C. M. Savage, Phys. Rev. A {\bf 56}, R1087
(1997).
\bibitem{9}  H. Steck, M. Naraschevski and H. Wallis, Phys. Rev. Lett
             {\bf 80}, 1 (1998).
\bibitem{10} M. Trippenbach, Y. Band, M. Edwards, M. Doery and
             P. S. Julienne, e-print cond-mat/9906033.
\bibitem{11} L. Deng, E. W. Hagley, J. Wen, M. Trippenbach, Y. Band,
             P. S. Julienne, J. E. Simsarian, K. Helmerson, S. L. Rolston,
             and W. D. Phillips, Nature {\bf 398}, 218 (1999).
\bibitem{12} M. G. Moore, O. Zobay and P. Meystre, e-print cond-mat/9902293.
\bibitem{13} G. Breitenbach, S. Schiller and J. Mlynek, Nature {\bf 387},
             471 (1997).
\bibitem{14} C. Sun, H. Zhan, Y. Miao, J. Li, Commun. Theor. Phys. {\bf 29},
             161 (1998).
\bibitem{15} D. F. Walls and G. J. Milburn, {\it Quantum Optics},
             (Springer-Verlag, Berlin, 1994).
\bibitem{16} F. Dalfovo, S. Giorgin, L. P. Pitaevskii and S. Stringari,
             Rev. Mod. Phys. {\bf 71}, 463 (1999).
\bibitem{17} V. Buzek, A. V. Barranco and P. L. Knight, Phys. Rev. A
             {\bf 45}, 6570 (1992).
 
\end{thebibliography}
\end{document}